\documentstyle[twocolumn]{esa_sp_l}

\begin{document}
%

\parindent 0pt
\parskip 10pt plus 1pt minus 1pt
\hoffset=-1.5truecm
\topmargin=-1.0cm
\textwidth 17.1truecm \columnsep 1truecm \columnseprule 0pt 

\input{epsf}

\def\exosat{{\it EXOSAT}}
\def\ginga{{\it Ginga}}
\def\heao{{\it HEAO-1}}
\def\rosat{{\it ROSAT}}
\def\asca{{\it ASCA}}

\def\ee{e$^\pm$}
\def\g{$\gamma$}
\def\sigmat{\sigma_{\rm T}}
\def\nh{N_{\rm H}}
\def\af{A_{\rm Fe}}
\def\taut{\tau_{\rm T}}
\def\ec{E_{\rm c}}
\def\ef{E_{\rm Fe}}
\def\sf{\sigma_{\rm Fe}}

\def\defeq{\equiv}
\def\prop{\propto}
\newbox\grsign \setbox\grsign=\hbox{$>$} \newdimen\grdimen \grdimen=\ht\grsign
\newbox\simlessbox \newbox\simgreatbox \newbox\simpropbox
\setbox\simgreatbox=\hbox{\raise.5ex\hbox{$>$}\llap
     {\lower.5ex\hbox{$\sim$}}}\ht1=\grdimen\dp1=0pt
\setbox\simlessbox=\hbox{\raise.5ex\hbox{$<$}\llap
     {\lower.5ex\hbox{$\sim$}}}\ht2=\grdimen\dp2=0pt
\setbox\simpropbox=\hbox{\raise.5ex\hbox{$\prop$}\llap
     {\lower.5ex\hbox{$\sim$}}}\ht2=\grdimen\dp2=0pt
\def\simgreat{\mathrel{\copy\simgreatbox}}
\def\simless{\mathrel{\copy\simlessbox}}
\def\simprop{\mathrel{\copy\simpropbox}}

\title{\bf X-RAYS AND GAMMA-RAYS FROM ACCRETION FLOWS ONTO BLACK HOLES\\
IN SEYFERTS AND X-RAY BINARIES}

\author{{\bf Andrzej A. Zdziarski}
\vspace{2mm} \\ 
 N. Copernicus Astronomical Center, Bartycka 18, 00-716 Warsaw, Poland
\vspace{2mm}\\
{\bf W. Neil Johnson} \vspace{2mm} \\
 E. O. Hulburt Center for Space Research, Naval Research Lab, Washington, 
DC 20375, USA
\vspace{2mm}\\
{\bf Juri Poutanen} \vspace{2mm}\\
 Uppsala Observatory, Box 515, S-75120 Uppsala, Sweden
\vspace{2mm}\\
{\bf Pawe\l\ Magdziarz and Marek Gierli\'nski} \vspace{2mm}\\
 Astronomical Observatory, Jagiellonian University, Orla 171, 30-244 
Cracow, Poland}
  
\maketitle

\begin{abstract}

We review observations and theoretical models of X-ray/\g-ray spectra of 
radio-quiet Seyfert galaxies and of Galactic black-hole candidates (in the 
hard spectral state). The observed spectra share all their basic components: 
an underlying power law, a Compton-reflection component with an Fe K$\alpha$ 
line, low-energy absorption by intervening cold matter, and a high-energy 
cutoff above $\sim 200$ keV. The X-ray energy spectral index, $\alpha$, is 
typically in the range $\sim 0.8$--1 in Seyfert spectra from \ginga, 
\exosat\/ and OSSE. The hard-state spectra of black-hole candidates Cyg X-1 
and GX 339-4 from simultaneous \ginga/OSSE observations have $\alpha\simeq 
0.6$--0.8. The Compton-reflection component corresponds to cold matter (e.g., 
inner or outer parts of an accretion disk) covering a solid angle of $\sim 
(0.4$--$1)\times 2\pi$ as seen from the X-ray source. The spectra are cut off 
in soft \g-rays above $\sim 200$ keV. The broad-band spectra of both Seyferts 
and black-hole sources are well fitted by Compton upscattering of soft photons 
in thermal plasmas. Our fits yield the thermal plasma temperature of $\sim 
100$ keV and the Thomson optical depth of $\tau\sim 1$.  A fraction of the 
luminosity emitted nonthermally appears to be small and it can be constrained 
to $\simless 15\%$ in the Seyfert galaxy NGC 4151. The spectra are cut off 
before 511 keV, which is strongly suggestive of a thermostatic role of \ee\ 
pair production in constraining the temperature and optical depth of the 
sources. The source geometry is compatible with a patchy corona above a cold 
disk in Seyferts, but not in Cyg X-1. In the latter, the relative weakness of 
reflection is compatible with reflection of emission of a hot inner disk from 
outer disk regions. 

 \vspace {5pt} 

  Keywords: 
galaxies: nuclei -- galaxies: Seyfert -- X-rays: galaxies -- 
gamma-rays: observations -- radiation mechanisms: thermal -- stars: individual: 
(Cyg X-1, GX 339-4)

\end{abstract}

\section{INTRODUCTION}

Seyfert galaxies are the first class of objects in which we are able to 
directly probe the gravitational potential near a black hole. This has 
been achieved thanks to the discovery by the \asca\/ observatory (Tanaka, 
Inoue \& Holt 1994) of broad Fe K$\alpha$ lines from several Seyfert 1s 
(Tanaka et al.\ 1995; Fabian et al.\ 1995; Iwasawa et al.\ 1996). The width of 
the lines, $\sim c/3$ or more, requires the lines originate in large part at 
radial distances of $6 GM/c^2$ or even less (Iwasawa et al.\ 1996). The peak 
energies of the narrow cores of the lines are close to 6.4 keV, which shows 
the lines are emitted by cold and dense media, with Fe being at most mildly 
ionized. 

The lines originate when X-rays irradiate cold matter. Photons with energies 
above the Fe K edge, $\simgreat 7$ keV, are able to ionize the K shell of an 
Fe atom or ion. Since the L shell is occupied (unless for He-like and H-like 
Fe ions), the ion becomes excited. The deexcitation proceeds in most cases 
through an L-to-K transition. The transition energy, $\geq 6.4$ keV, goes into 
either an ejection of an outer shell (Auger) electron, or into emission of a 
K$\alpha$ photon. The probability of the latter process is about 50\% (e.g., 
George \& Fabian 1991). The resulting fluorescent K$\alpha$ line is broadened 
by the Doppler effect and gravitationally. 

A very important process associated with the fluorescent formation of Fe
K$\alpha$ lines is Compton reflection (White, Lightman \& Zdziarski 1988; 
Lightman \& White 1988). In this process, X-rays and \g-rays (hereafter X\g) 
irradiating the cold matter get Compton-scattered (in one or more scattering 
events) back to the observer. Compton reflection of a power-law spectrum 
typical for Seyferts forms a characteristic continuum peaked (in $\nu F_\nu$) 
around 30 keV. Below the peak, bound-free absorption becomes important and 
incident photons are likely to get absorbed. The bound-free cross section of 
cosmic-composition matter increases (except at ionizatoin edge energies) with 
decreasing energy, and it becomes larger than the Thomson cross section below 
$\sim 10$ keV. Above the peak, Klein-Nishina effects become important. Namely, 
a photon loses a substantial part of its energy in a scattering, which
occurs preferentially forward, and its cross section decreases with energy. 
These effect results in a steep cutoff of the reflected spectrum above $\sim 
100$ keV. The presence of a Compton-reflected spectral component has been 
discovered in the spectra of Seyferts by \ginga\/ (Pounds et al.\ 1990; Nandra 
\& Pounds 1994). The relative normalization of the reflected component with 
respect to the direct one in the spectra of Seyfert 1s corresponds to the solid 
angle of the reflector close to $2\pi$ (Nandra \& Pounds 1994; Zdziarski et 
al.\ 1995; Gondek et al.\ 1996). This strongly suggests that the reflector is 
a cold accretion disk. This conclusion is further supported by the fits to the 
K$\alpha$ line profiles by emission of relativistic disks (extending to the 
minimum stable orbit, e.g., Tanaka et al.\ 1995) and the large equivalent 
widths of the K$\alpha$ lines, EW$\sim 200$ eV on average (Nandra et al.\ 
1996).  

Thus, the results of \asca\/ and \ginga\/ lead to a picture with the X\g\ 
emission in Seyfert 1s typically taking place above the surface of an 
accretion disk. As discussed above, we understand relatively well the 
reflected, secondary, component of the emission. On the other hand, the origin 
of the incident, {\it primary}, radiation is much less understood. In order to 
understand its origin, we need to study broad-band, X\g\ spectra of Seyferts.
In X-rays, the intrinsic spectra turn out to be simple power laws (see \S 2 
below), which only weakly constrain their origin. A key observable needed to 
understand the origin of the primary spectra is the spectral shape at high 
energies, in particular any spectral break or cutoff, or a spectral feature. 
This key information can be provided by soft \g-ray observations by OSSE 
(Johnson et al.\ 1993) aboard {\it Compton Gamma Ray Observatory}, as well as 
future soft \g-ray experiments, e.g., {\it INTEGRAL\/} (Winkler 1994). In this 
work, we review the available X\g\ spectra from Seyferts and discuss resulting 
constraints on physical processes in the radiation sources (see also reviews 
by Svensson 1996; Zdziarski et al.\ 1996). 

We also discuss the spectral properties of Galactic black-hole candidates in 
the hard (so-called `low') state (see Nowak 1995). We consider spectra of GX 
339-4 and Cyg X-1. The spectra are very similar to those of Seyferts and 
consist of an underlying power-law, a Compton reflection component, and a 
high-energy cutoff at $\sim 200$ keV. This remarkable similarity between 
Seyfert and stellar-mass black-hole sources appears to reflect the shared 
origin of the emission from an accretion flow onto a black hole. We find that 
the intrinsic power laws with the cutoff in both classes of objects are well 
modeled by thermal Compton upscattering in mildly relativistic plasmas of 
moderate Thomson optical thickness ($kT\sim 100$ keV, $\tau\sim 1$).

\section{SPECTRA OF IC 4329A AND OF AVERAGE SEYFERT SAMPLES}

As yet, relatively few broad-band, X\g\ spectra of Seyferts are available. 
Those include the spectrum of IC 4329A (a Seyfert 1) from \rosat, \ginga, and 
OSSE (Madejski et al.\ 1995). The spectrum (fitted as in Zdziarski et al.\ 
1996) is shown in Figure 1a. It consists of an underlying power law with the 
energy spectral index of $\alpha= 0.94\pm 0.03$ and a reflection component 
corresponding to the solid angle of $\Omega/2\pi= 0.74^{+0.17}_{-0.16}$ 
(assuming the viewing anle of $30^\circ$ in angle-dependent reflection Green's 
functions of Magdziarz \& Zdziarski 1995 and using the elemental abundances of 
Anders \& Grevesse 1989). There is a cutoff in the spectrum above $\sim 200$ 
keV; when the spectrum is fitted with an e-folded power law, the e-folding 
energy is $\ec=0.3^{+0.2}_{-0.1}$ MeV. (We use {\sc xspec}, see Arnaud 1996, 
for spectral fitting, and the quoted errors are for 90\% confidence limit 
based on a $\Delta \chi^2=2.7$ criterion.)

\begin{figure}
\begin{center}
\leavevmode
\epsfxsize=8.1cm \epsfbox{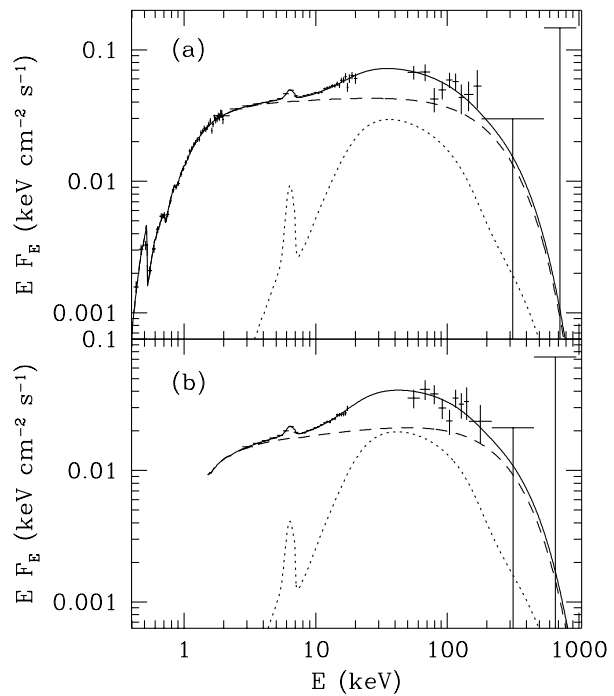}  
\end{center}
\caption{
The broad-band X\g\ spectra of Seyfert 1s. {\it (a)} The \rosat/\ginga/OSSE 
spectrum of IC 4329A. {\it (b)} The average spectrum ({\it crosses}) of 5 
radio-quiet Seyfert 1s detected by both OSSE and \ginga. The upper limits here 
and in figures below are 2-$\sigma$.  The dashed curves represent the best-fit 
thermal Comptonization spectra of PS96. The dotted curves represent the 
absorbed reflected component including the fluorescent Fe K$\alpha$ line. The 
solid curves give the sum.} 
 \end{figure} 

A similar cutoff above $\sim 200$ keV is seen in the average spectrum of 5 
radio-quiet Seyfert 1s observed 25 times by \ginga\/ and 14 times by OSSE 
(Gondek et al.\ 1996), shown in Figure 1b. The parameters of the fit with an 
e-folded power law are $\alpha=0.90\pm 0.05$, $\Omega/2\pi=0.76\pm 0.15$, and 
$\ec=0.7^{+2.0}_{-0.3}$ MeV (assuming the abundances of Anders \& Ebihara 
1982). A very similar average spectrum is also obtained for 7 radio-quiet 
Seyfert 1s observed 41 times by \exosat\/ and 18 times by OSSE (Gondek et al.\ 
1996). 

A major question for our understanding of the AGN phenomenon is the origin of 
the underlying cut-off power law spectrum. Models with synchrotron emission 
predict simultaneous IR/X-ray variability, which is not observed. Also, the 
cutoff above 200 keV would then require a fine-tuned sharp cutoff in the 
distribution of the emitting relativistic electrons. The most likely physical 
process responsible for the underlying spectra appears to be Compton 
scattering in a thermal plasma. The formation of power-law spectra in this 
model is due to superposition of relatively narrow scattering profiles from 
subsequent orders of scattering (e.g., Svensson 1996). The spectral index is a 
function of the optical depth and temperature of the plasma, and there is 
high-energy cutoff in the spectrum at energies above a few times $kT$. The 
plasma in Seyfert 1s is {\it not\/} optically thick (Zdziarski et al.\ 1994), 
which precludes the use of formulae of Sunyaev \& Titarchuk (1980). Until 
recently, there have been no ready-to-use, accurate, formulae for spectra of 
$\tau\simless 1$ plasmas (as discussed in Poutanen \& Svensson 1996, hereafter 
PS96). Instead, power-laws with exponential cutoffs have been commonly used 
(see above). 

Here, we use the code of PS96 for thermal Comptonization in active regions 
above the surface of a disk. The active regions are assumed to have the form 
of hemispheres (with the radial optical depth equal $\tau$) on the disk 
surface. The code gives highly accurate results, which agree with those 
obtained with a Monte Carlo method (Stern et al.\ 1995; PS96). Here, we use 
this code to fit IC 4329A (see also Poutanen, Svensson \& Stern 1997) and the 
average \ginga/OSSE spectrum of Seyfert 1s (Gondek et al.\ 1996). Our 
preliminary fits give $\tau=1.3^{+0.4}_{-0.2}$,  $kT= 100^{+10}_{-30}$ keV, 
and $\tau=1.0^{+0.4}_{-0.2}$ and $kT= 130^{+30}_{-40}$ keV, respectively. The 
model spectra are shown by the curves in Figures 1a, b. 

The statistics of the OSSE data of the two above spectra is relatively 
limited, and thus the cutoff is not very well constrained. Therefore, we also 
consider the co-added spectrum of all Seyfert 1s (except NGC 4151) detected by 
OSSE through 1995 (McNaron-Brown et al.\ 1997). The sample consists of 29 
observations of 12 Seyfert 1s. NGC 4151, whose flux in the OSSE range is an 
order of magnitude higher than that of other Seyfert 1s, is excluded in order 
to avoid the dominance of the spectrum by a single object. That spectrum, 
shown in Figure 2a, has the shape indistinguishable from that of the co-added 
OSSE spectra of Gondek et al.\ (1996). By this we hereafter mean that the two 
spectra can be fitted with the same model with $\Delta\chi^2 < 2.7$ with 
respect to the sum of $\chi^2$ for the two individual fits. The present OSSE 
spectrum shows a clear high-energy cutoff above $\sim 150$ keV. However, the 
OSSE spectrum starts at 50 keV only and it does not allow to determine the X-
ray spectral index and the normalization of the reflection component, which 
quantities are both are important for determining the plasma temperature. 
Therefore, while fitting the average OSSE spectrum we constrain the X-ray 
spectrum at the best-fit values obtained for the \ginga/OSSE average spectrum 
($\alpha=0.90\pm 0.05$, $\Omega/2\pi = 0.76\pm 0.15$). Our preliminary fit 
results are again very similar to those above, $\tau= 1.2^{+0.4}_{-0.3}$ and 
$kT= 110^{+40}_{-30}$ (shown by the curves in Figure 2a). 

\begin{figure}
\begin{center}
\leavevmode
\epsfxsize=8.1cm \epsfbox{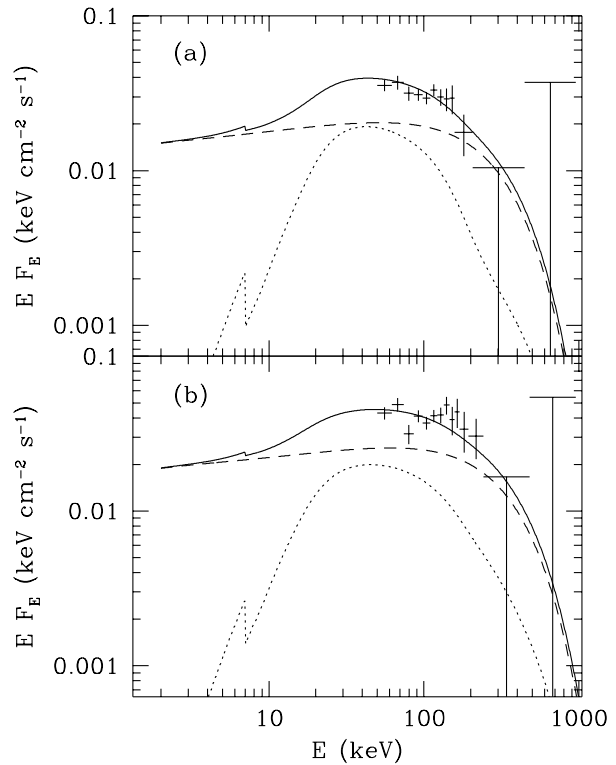}
\end{center}
\caption{
The average spectra ({\it crosses}) of radio-quiet Seyferts detected by 
OSSE (McNaron-Brown et al.\ 1997). The dashed curves represent the best-fit 
thermal Comptonization spectra of PS96 (without showing absorption at soft 
X-rays). The dotted curves represent the reflected component. The solid curves 
give the sum. {\it (a)} The Seyfert 1 sample (excluding NGC 4151). {\it (b)} 
The Seyfert 2 sample. } 
 \end{figure}

We compare now the average OSSE spectrum of Seyfert 1s with that of Seyfert 2s 
detected by OSSE (8 objects observed 19 times through 1995; McNaron-Brown et 
al.\ 1997). We find that the spectrum, shown in Figure 2b, is statistically 
indistinguishable from that of Seyfert 1s above, and it again shows a cutoff 
above $\sim 200$ keV. When we fit it with the best-fit constraints at X-rays 
as obtained by Gondek et al.\ (1996) for Seyfert 1s (except that the disk 
viewing angle, important for reflection, is fixed at $60^\circ$, compatible
with the AGN unified model, Antonucci 1993), we obtain $\tau= 1.1^{+0.2}_{-
0.2}$ and $kT= 110^{+20}_{-20}$ (corresponding to the curves in Figure 2b). 
The fact that these fits give results virtually identical to those of Seyfert 
1s strongly supports the unified AGN model, according to which Seyfert 1s and 
2s differ only by orientation with respect to the observer (Antonucci \& 
Miller 1985; Antonucci 1993). It also shows that the characteristic column 
densities of the obscuring torii in the considered sample are such that 
photons at 50 keV are not strongly absorbed, which corresponds to $\nh 
\simless 10^{24}$ cm$^{-2}$.

The plasma inferred here to exist in Seyferts has $\tau\sim 1$ and $kT\sim 
100$ keV, i.e., it has larger optical depth and lower temperature from the 
parameters considered before, e.g., by Zdziarski et al.\ (1994) and Stern et 
al.\ (1995). The former authors fitted the spectrum of IC 4329A by an e-folded 
power law with reflection, and only then found the plasma parameters 
corresponding to the best fit. However, the actual shape of the high-energy 
cutoff from thermal Comptonization is not described by an e-folded power law, 
and thus that procedure led to an underestimate of $\tau$ and an overestimate 
of $kT$. On the other hand, Stern et al.\ (1995) considered models with hot, 
pure \ee-pair, plasma regions attached to the disk surface and compared the 
results with the distribution of X-ray spectral indices in Seyfert 1s of 
Nandra \& Pounds (1994). This implied that $\tau\simless 0.3$ correspond to 
hot plasmas in Seyferts. This disagrees with the present fits, which implies 
that some of the assumption of Stern et al.\ are not satisfied. It can be 
either that the active regions are located at some height {\it above\/} the 
disk surface (Svensson 1996), or the plasma is not pure \ee\ pairs, or both.

\section{NGC 4151 AND LIMIT ON NONTHERMAL PROCESSES}

As mentioned above, the brightest radio-quiet Seyfert is NGC 4151, observed by 
OSSE 10 times during 1991-96 (Johnson et al.\ 1997). Its X-ray spectrum is 
strongly absorbed by $\nh \sim 10^{23}$ cm$^{-2}$. Therefore, it is difficult 
to determine the form of the intrinsic X-ray spectrum. X\g\ observations 
before 1991 show hard intrinsic X-ray spectra, with $\alpha\sim 0.3$--0.8 and 
no detectable Compton-reflection component (e.g.\ Yaqoob et al.\ 1993). On the 
other hand, contemporaneous X\g\ observations in 1991--93 June show the object 
in a soft state, with $\alpha \sim 0.80\pm 0.05$ (when fitted with thermal 
Comptonization, Zdziarski, Johnson \& Magdziarz 1996, hereafter ZJM96). This 
X-ray spectral index is within the 1-$\sigma$ range of those observed on 
average in Seyfert 1s, $\langle \alpha \rangle=0.95$, $\sigma=0.15$ (Nandra \& 
Pounds 1994). Furthermore, the \ginga/OSSE observation of 1991 June/July shows 
the presence of a Compton-reflection component with $R\simeq 0.4\pm 0.2$, see 
Figure 3a (ZJM96). Also, the OSSE spectra of NGC 4151 are statistically not 
distinguishable from the average OSSE spectrum of Seyfert 1s (ZJM96). The 
corresponding plasma parameters determined by ZJM96 are $\tau\sim 1.3$, 
$kT\sim 60$ keV. Thus, the only characteristic distinguishing the soft state 
of NGC 4151 from other Seyfert 1s appears to be a large absorbing column. 

The simultaneous \asca/OSSE spectrum of 1993 May (ZJM96) does not allow a 
determination of the strength of Compton reflection, which leads to some 
ambiguity in determining the intrinsic spectrum. If the same amount of 
reflection as for the \ginga/OSSE observation is assumed, the intrinsic 
spectrum is the same within statistical uncertainties as the \ginga/OSSE 
spectrum, see Figure 3b (ZJM96).

\begin{figure}
\begin{center}
\leavevmode
\epsfxsize=8.1cm \epsfbox{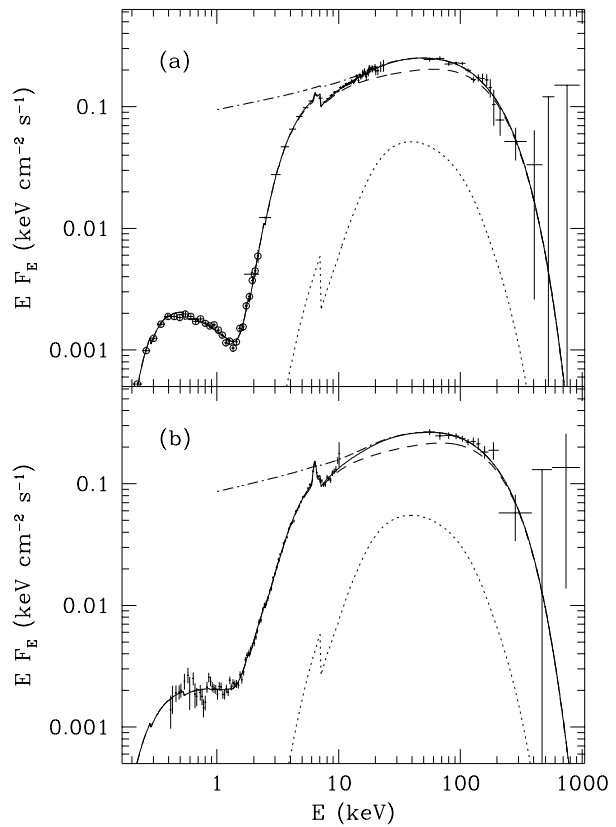}
\end{center}
 \caption{
The X\g\ spectra of NGC 4151 observed by {\it (a)} \rosat, \ginga\/ and OSSE 
in 1991 June/July and {\it (b)} by \asca and OSSE in 1993 May (ZJM96). The 
dot-dashed curves show the unabsorbed spectrum (without the Fe K$\alpha$ 
line and the separate soft X-ray component, dominant below 1 keV). The dashed 
curves represent the best-fit thermal Comptonization spectra of ZJM96. The 
dotted curves represent the absorbed reflected component.
The solid curves give the total model spectra. } 
 \end{figure}

OSSE observations of NGC 4151 yield spectra relatively constant in shape, and 
with the normalization varying within a factor of $\sim 2$ (ZJM96; Johnson et 
al.\ 1997). Thus, the co-added spectrum is representative of time-resolved 
spectra but it gives much better statistics at high energies. The spectrum is 
shown in Figure 4. We again obtain that the average NGC 4151 spectrum is 
statistically not distinguishable from that of the average Seyfert-1 spectrum. 
This again argues for NGC 4151 being a relatively average Seyfert 1 rather 
than a peculiar object. The plasma parameters for the average spectrum of NGC 
4151 fitted with the thermal Comptonization model of ZJM96, $\tau \sim 1.3$, 
$kT \sim 70$ keV, are similar to those in other Seyfert 1s.

\begin{figure}
\begin{center}
\leavevmode
\epsfxsize=8.1cm \epsfbox{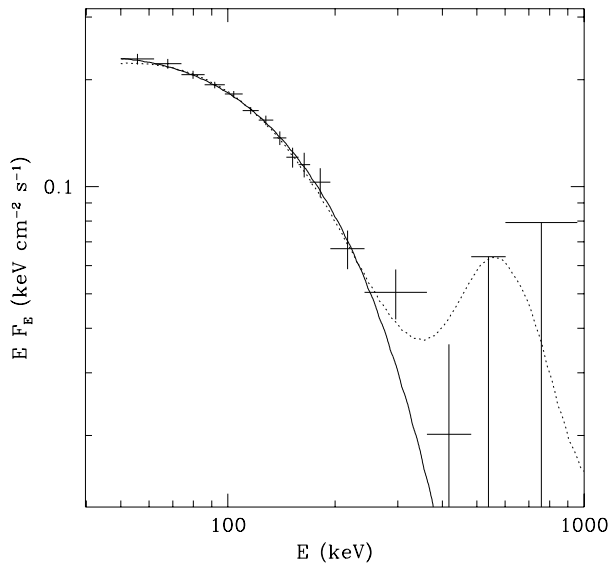}
\end{center}
 \caption{
The average OSSE spectrum of NGC 4151 (Johnson et al.\ 1997). The solid curve 
gives the best-fit thermal Comptonization model. The dashed curve corresponds 
to the spectrum with the strongest allowed fraction, 17\%, of nonthermal 
processes in the source. See text. } 
 \end{figure}

We see in Figure 4 that the average OSSE spectrum of NGC 4151 shows a cutoff 
well-fitted by the thermal Comptonization model. We can use this spectrum to 
constrain the presence of nonthermal processes in the X\g\ source in NGC 4151. 
We consider a model in which a fraction of the available power is used to 
accelerate selected electrons (or \ee\ pairs) to relativistic energies, and 
the rest is distributed approximately equally among the remaining electrons, 
i.e., it heats the plasma. Such a model for NGC 4151 was proposed by Zdziarski, 
Lightman \& Macio{\l}ek-Nied\'zwiecki (1993). The presence of nonthermal 
acceleration gives rise to a tail on top of the thermal spectrum. This tail is 
due to both Compton scattering by the relativistic electrons as well as due to 
annihilation of \ee\ pairs from the resulting pair cascade (see Svensson 1987; 
Lightman \& Zdziarski 1987). We repeat the calculations of Zdziarski et 
al.\ (1993) for the average OSSE spectrum of NGC 4151 (Johnson et al.\ 1997). 
We find that the presence of a nonthermal tail does not improve the fit to the 
spectrum. The 90\% confidence limit on the nonthermal fraction is 0.17. The 
spectrum corresponding to this limit is shown in Figure 4 in dotted line.

\section{GALACTIC BLACK-HOLE CANDIDATES IN THE HARD STATE}

It is very intriguing that some Galactic black-hole candidates observed in the 
hard (so-called `low') state have high-energy spectra virtually identical to 
those of Seyfert 1s. One example is the hard-state spectrum of GX 339-4 
observed simultaneously by \ginga\/ and OSSE in 1991 September (Ueda et al.\ 
1994; Grabelsky et al.\ 1995). The spectrum, re-analyzed and fitted 
with the thermal Comptonization model of ZJM96, is shown in Figure 5. The 
spectral parameters are virtually identical to those of NGC 4151, with 
$kT\simeq 70$ keV, $\alpha=0.8$ (corresponding to $\tau\sim 1.2$), and a 
Compton reflection component corresponding to $\Omega/2\pi\simeq 0.5$.

\begin{figure} 
\begin{center} 
\leavevmode 
\epsfxsize=8.1cm \epsfbox{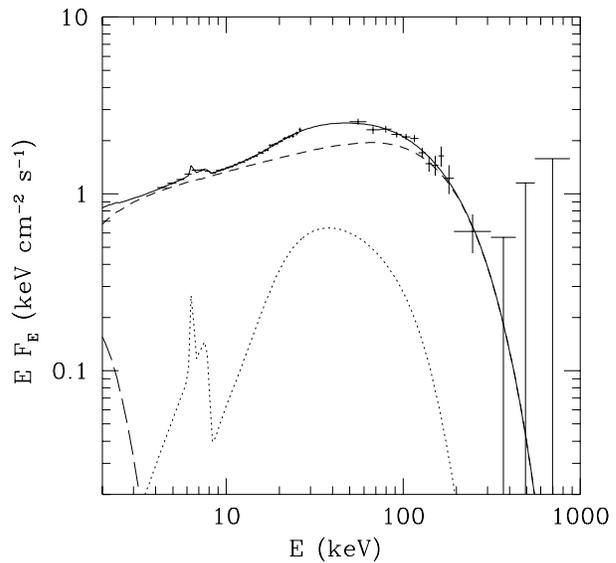} 
\end{center} 
\caption{An X\g\ spectrum of GX 339-4 observed simultaneously by \ginga\/ and 
OSSE and fitted by the thermal Comptonization model of ZJM96 (dashed curve) 
and Compton reflection (dotted curve). The long dashes correspond to a soft 
component due to blackbody disk emission. The solid curve gives the sum. 
 } 
\end{figure}

The archetypical black-hole candidate Cyg X-1 was observed four times by 
\ginga\/ and OSSE simultaneously in 1991 June (Gierli\'nski et al.\ 1996; 
1997).  The spectra correspond to the hard (`low') state of this source. They 
show $\alpha\simeq 0.6$, a reflection component, and a high-energy cutoff 
above $\sim 150$ keV similar to that of AGNs. However, that cutoff is sharper 
than one possible to obtain with a uniform thermal plasma. Gierli\'nski et 
al.\ (1997) fitted that spectrum with the sum of Comptonization spectra in two 
plasma clouds with $\tau\sim 1$, $kT\simeq 140$ keV and $\tau\simeq 7$ and 
$kT\simeq 50$ keV, respectively. The first component has the parameters very 
similar to those obtained above for Seyeferts. It may well be that high-energy 
cutoffs in Seyferts are also complex, but the limited statistics of Seyfert 
spectra precludes constraining fits more complex than those corresponding to 
uniform plasma in the vicinity of cold matter. The similarity between the X\g\ 
spectra of Seyferts and Galactic black hole sources is, in fact, expected due 
to scale-free character of both disk accretion and two-body processes.

\begin{figure} 
\begin{center} 
\leavevmode 
\epsfxsize=8.1cm \epsfbox{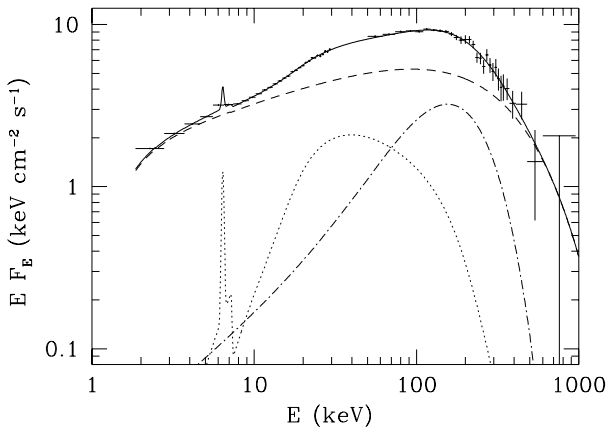} 
\end{center} 
\caption{An X\g\ spectrum of Cyg X-1 observed simultaneously by 
\ginga\/ and OSSE and fitted by the sum of a power law with an exponential 
cutoff (dashed curve), Compton reflection (dotted curve), and an additional 
thermal-Compton component from an optically thick plasma (dot-dashed curve). 
The solid curve gives the sum. From Gierli\'nski et al.\ (1997).
 } 
\end{figure}

\section{GEOMETRY}

We can constrain the geometry of thermal Comptonization sources by considering 
energy balance. There are two main conditions. First, a cloud of hot plasma 
will upscatter incident soft, seed, photons into the X\g\ range, in which 
process the incident soft photon flux will get amplified by an amplification 
factor, $A$, dependent on $\tau$ and $kT$. This can be determined by spectral 
fitting, and for sources with hard spectra (with $\alpha<1$), $A\gg 1$. Thus, 
 $$L_{{\rm X}\gamma}= A(\tau, kT) L_{\rm soft}^{\rm incident}.$$
 Second, the X\g\ flux irradiating cold matter will get reprocessed and 
reradiated mostly as soft photons (the integrated albedo for Compton 
reflection is small, $a\sim 0.1$, e.g., Magdziarz \& Zdziarski 1995). For an 
isotropic X\g\ source, 
 $$L_{\rm soft}^{\rm reprocessed} =(\Omega/2\pi)(1-a) L_{{\rm X}\gamma}, $$ 
 (where $\Omega$ is the solid angle subtended by the reflecting and 
reprocessing medium, as in Sections above). From observations, $\Omega\sim 
2\pi$ (see above). Thus, $L_{\rm soft}^{\rm reprocessed} \gg L_{\rm 
soft}^{\rm incident}$ when $A\gg 1$. This rules out a homogeneous corona 
geometry (proposed by Haardt \& Maraschi 1993) for objects with hard spectra 
($\alpha\simless 1$). 

Note that the above result is different from that of Haardt, Maraschi \& 
Ghisellini (1994), who pointed out that $L_{\rm UV}\gg L_{{\rm X}\gamma}$ 
often observed in Seyferts (Walter \& Fink 1993) also rules out the 
dissipative, homogeneous, corona model, in which the two quantities are 
predicted to be similar. These observations require that dissipation
takes place in the cold matter (presumably an accretion disk), not only in the 
corona (as assumed in Haardt \& Maraschi 1993).  

A geometry that can solve both of the problems above is a {\it patchy\/} 
corona, in which the hot plasma forms small active regions above the disk 
surface. The ratio of the size of an active region to its average height above 
the disk determines what fraction of the reprocessed photons (with $L_{\rm 
soft}^{\rm reprocessed}$) returns to the active region, and thus becomes 
$L_{\rm soft}^{\rm incident}$ (Stern et al.\ 1995; Svensson 1996). For the 
characteristic parameters of compact objects found here ($\tau\sim 1$ and 
$kT\sim 100$ keV), active regions located above the disk surface at a height 
similar to the size of the active region satisfy the energy balance (Fig.\ 2 
in Svensson 1996). On the other hand, the UV luminosities in Seyferts are 
often much larger than the X\g\ ones, which requires that the active regions 
cover a relatively small fraction of the disk surface. That fraction is 
determined by both the luminosity ratio and the ratio of the dissipation rate 
in the corona to that in the disk (Haardt et al.\ 1994). 

Thus, patchy coronae provide a good model for sources with $\Omega\sim 2\pi$ 
(as determined by the Compton reflection fits). The reflecting (and 
reprocessing) disks and their patchy coronae in Seyferts often extend to the 
minimum stable orbit, as indicated by the observations of broad Fe K$\alpha$ 
lines by \asca\/ (e.g., Fabian et al.\ 1995). If $\Omega\ll 2\pi$ (as seems to 
be the case in Cyg X-1, see above), the hot sources cannot be located above a 
disk. A possible geometry is then a hot inner disk and a cold outer disk (as 
discussed for Cyg X-1 by Gierli\'nski et al.\ 1997). This is also compatible 
with the narrowness of the Fe K$\alpha$ line in Cyg X-1 (Ebisawa et al.\ 
1996).

\section{ELECTRON-POSITRON PAIRS}

The next issue we consider is the presence of \ee\ pairs in the sources. The 
role of pair production is qualitatively different in sources with and without 
nonthermal processes. When nonthermal processes dominate, Compton upscattering 
of seed soft photons by nonthermal, relativistic, electrons can easily produce 
photons well above the threshold for \ee\ pair production, 511 keV. The 
photons with energies $>511$ keV can be absorbed in photon-photon 
interactions, which leads to a prediction of a cutoff in the spectrum above 
511 keV. The produced pairs can be relativistic as well and further Compton  
upscatter the seed photons to energies above 511 keV, which gives rise to a 
pair cascade (Svensson 1987; Lightman \& Zdziarski 1987). The pairs 
annihilate, however, only after loosing most of their energy and having joint 
the low-energy, thermal, electron distribution. Therefore, the annihilating 
pairs have typically energies much lower than the energies of electrons that 
lead to the pair production process. Thus, a common prediction of nonthermal 
models is the presence of a strong and \ee\ pair annihilation feature around 
511 keV (e.g., Svensson 1987; Lightman \& Zdziarski 1987). Such features have 
not been observed in AGNs. The average OSSE spectrum of NGC 4151 constrains 
the power channeled into nonthermal processes (which could lead to a pair 
annihilation feature, see Figure 4), to less than 17\% (Section 3). Also, no 
annihilation feature is seen in the spectra of the Galactic black-hole source 
Cyg X-1 (Phlips et al.\ 1996). Therefore, acceleration of electrons to 
relativistic energies appears to play at most a minor role in black hole 
accretion flows. 

On the other hand, good fits to the spectra of compacts objects are obtained 
with thermal models (see Sections 2-4). In thermal sources, pairs and 
electrons have a Maxwellian distribution. Compton scattering of seed photons 
by the pairs and electrons determines the X\g\ spectrum, which may extend 
above 511 keV. The pair production rate follows then from the shape of the 
spectrum. The pair annihilation rate depends mostly on the optical depth of 
the source. In pair equilibrium, the two quantities are equal, which condition 
determines the compactness parameter, 
 $$\ell \equiv l\sigma_{\rm T}/r m_e c^3$$
 (where $l$ and $r$ are the luminosity and size, respectively, of an active 
region, and $\sigma_{\rm T}$ is the Thomson cross section), at which the 
source is made entirely of pairs in steady state (Svensson 1984). This 
represents the upper limit to the actual compactness, which can be is lower if 
there are some ionization electrons in addition to \ee\ pairs. The compactness 
requires for pair-domination is also lower if some fraction of the luminosity 
is channeled into nonthermal processes (e.g., ZJM96). The plasma parameters 
inferred for most of the sources considered in this review are $\tau\sim 1$ 
and $kT\sim 100$ keV, for which the maximum $\ell\simeq 300$ (see Fig.\ 3 in 
Svensson 1996). Such a compactness is possible for sources accreting at a 
fraction of the Eddington limit. 

Note that even if a thermal source is made entirely of \ee\ pairs, no distinct 
pair annihilation feature around 511 keV is visible in the spectrum 
(Macio{\l}ek-Nied\'zwiecki, Zdziarski \& Coppi 1995). Thus, the lack of 
detection of annihilation lines in the spectra of compact objects is still 
compatible with their composition of thermal \ee\ pairs. 

The way pair production constrains source parameters in thermal sources is 
{\it not\/} by pair absorption (important above 511 keV in nonthermal 
sources). Rather, pair production leads to an increase of the number of 
particles in the source. If the total power supplied to the particles is 
fixed, this leads to a decrease of the average energy per particle, or the 
temperature. The lowered temperature leads to fewer photons upscattered above 
511 keV, which in turn lowers the pair production rate. The fact that the X\g\ 
spectra of compact objects are cut off close to 511 keV (at $\sim 200$ keV, 
see above), strongly suggests that this pair thermostat operates in Seyferts 
and black-hole binaries. The typical plasma temperature of 100 keV found here 
is such that the sources are just at the onset of copious pair production. 

\section{CONCLUSIONS}

Our main new result is that the plasma parameters in Seyferts and in Galactic 
black-hole candidates in the hard state are $\tau\sim 1$ and $kT\sim 100$ keV. 
This has been obtained by direct fitting the available X\g\ spectra with 
thermal Comptonization models. The spectra are cut off at energies below the 
electron rest energy, which indicates that \ee\ pair production is an 
important process. Nonthermal processes appear to play at most a minor role. 

The presence of Compton-reflection components shows that there is cold matter 
subtending a substantial solid angle as seen from the X\g\ source. 
In Seyferts, a patchy corona above a cold accretion disk is consistent with 
the data. On the other hand, this geometry is not compatible with the spectrum 
of Cyg X-1, in which the cold matter subtends a relatively small solid angle, 
consistent with reflection from an outer disk. This is also consistent the 
lack of detection of broad Fe K$\alpha$ lines in black-hole sources, which 
argues against the presence of cold matter close to the innermost stable 
orbit.

\section*{ACKNOWLEDGMENTS}

This research has been supported in part by NASA grants and contracts and the 
Polish KBN grants 2P03D01008, 2P03D01410, 2P03C00511p01, and 2P03C0511p4.
We thank Bronek Rudak for help with fitting GX 339-4.


\begin{thebibliography}{}

\bibitem[]{}
Anders E., Ebihara M., 1982, Geochim.\ Cosmochim.\ Acta 46, 2363

\bibitem[]{}
Anders E., Grevesse N., 1989, Geochim.\ Cosmochim.\ Acta 53, 197

\bibitem[]{}
Antonucci R. R. J., 1993, ARAA, 31, 473

\bibitem[]{}
Antonucci R. R. J., Miller J. S., 1985, ApJ, 297, 621

\bibitem[]{}
Arnaud K. A., 1996, in: Jacoby G. H., Barnes J. (eds.) Astronomical Data 
Analysis Software and Systems V. ASP Conf.\ Series Vol.\ 101, San Francisco, 
p.\ 17


\bibitem[]{}
Ebisawa, K., Ueda Y., Inoue H., Tanaka Y., White N. E., 1996, ApJ, 467, 419


\bibitem[]{}
Fabian A. C., Nandra K., Reynolds C. S., Brandt W. N., Otani C., Tanaka Y., 
Inoue H., Iwasawa K., 1995, MNRAS, 277, L11 


\bibitem[]{}
George I. M., Fabian A. C., 1991, MNRAS, 249, 352


\bibitem[]{}
Gierli\'nski M., Zdziarski A. A., Johnson W. N., et al.,
1996, in: Zimmermann H. U., Tr\"umper J., Yorke H. (eds.) 
MPE Report 263, R\"ontgenstrahlung from the Universe, p.\ 139 

\bibitem[]{}
Gierli\'nski M., Zdziarski A. A, Done C., Johnson W. N., Ebisawa K., Ueda Y., 
Haardt F., Phlips B. F., 1997, MNRAS, submitted 


\bibitem[]{} Gondek D., Zdziarski A. A., Johnson W. N., George I. M., 
McNaron-Brown K., Magdziarz P., Smith D., Gruber D. E., 1996, MNRAS, 282, 646 

\bibitem[]{}                                                   
Grabelsky D. A., Matz S. M., Purcell W. R., et al., 1995, ApJ, 441, 800

\bibitem[]{}
Haardt F., Maraschi L., 1993, ApJ, 413, 507

\bibitem[]{} 
Haardt F., Maraschi L., Ghisellini G., 1994, ApJ, 432, L95

\bibitem[]{} 
Iwasawa K., et al., 1996, MNRAS, 282, 1038

\bibitem[]{} 
Johnson W. N., et al., 1993, ApJS, 86, 693

\bibitem[]{} 
Johnson W. N., McNaron-Brown K., Kurfess J. D., Zdziarski A. A., Magdziarz 
P., Gehrels N., Reichert G. A., 1997, ApJ, submitted


\bibitem[]{} 
Lightman A. P., White T. R., 1988, ApJ, 335, 57  

\bibitem[]{}
Lightman A. P., Zdziarski A. A., 1987, ApJ, 319, 643 

\bibitem[]{}
Macio{\l}ek-Nied\'zwiecki A., Zdziarski A. A., Coppi P. S., 1995, 
MNRAS, 276, 273

\bibitem[]{}
Madejski G. M., Zdziarski A. A., Turner T. J., et al., 1995, ApJ, 438, 672 

\bibitem[]{} 
Magdziarz P., Zdziarski A. A., 1995, MNRAS, 273, 837


\bibitem[]{}
McNaron-Brown, K., et al., 1997, in preparation

\bibitem[]{} 
Nandra K., George I. M., Mushotzky R. F., Turner T. J., Yaqoob Y., 1996, ApJ, 
in press

\bibitem[]{} 
Nandra K., Pounds K., 1994, MNRAS, 268, 405 

\bibitem[]{}
Nowak M. A., 1995, PASP, 718, 1207

\bibitem[]{}
Phlips B. F., Jung G. V., Leising M. D., et al., 1996, ApJ, 465, 907

\bibitem[]{}
Pounds K. A., Nandra K., Stewart G. C., George I. M., Fabian A. C., 
1990, Nature, 344, 132

\bibitem[]{}
Poutanen J., Svensson R., 1996, ApJ, in press (PS96)

\bibitem[]{}
Poutanen J., Svensson R., Stern B., 1997, in:
Winkler C., Courvousier T., Durouchoux P. (eds.) ESA SP-382, 
The Transparent Universe, in press



\bibitem[]{} 
Stern B. E., Poutanen J., Svensson R., Sikora M., 1995, ApJ, 449, L13

\bibitem[]{} 
Sunyaev R. A., Titarchuk L. G., 1980, A\&A, 86, 121


\bibitem[]{} 
Svensson R., 1984, MNRAS, 209, 175

\bibitem[]{} 
Svensson R., 1987, MNRAS, 227, 403

\bibitem[]{} 
Svensson R., 1996, A\&AS, 120, in press


\bibitem[]{}
Tanaka Y., Inoue H., Holt S. S., 1994, PASJ, 46, L37

\bibitem[]{}
Tanaka Y., et al., 1995, Nature, 375, 659

\bibitem[]{}
Ueda Y., Ebisawa K., Done C., 1994, PASJ, 46, 107

\bibitem[]{}
Walter R., Fink H. H., 1993, A\&A, 274, 105


\bibitem[]{} 
White T. R., Lightman A. P., Zdziarski A. A., 1988, ApJ, 331, 939

\bibitem[]{} 
Winkler C., 1994, ApJS, 92, 327


\bibitem[]{}
Zdziarski A. A., Fabian A. C., Nandra K., Celotti A., Rees M. J., Done C., 
Coppi P. S., Madejski G. M., 1994, MNRAS, 269, L55 

\bibitem[]{}
Zdziarski A. A., Gierli\'nski M., Gondek D., Magdziarz P., 
1996, A\&AS, 120, in press

\bibitem[]{}
Zdziarski A. A., Johnson W. N., Done C., Smith D., McNaron-Brown K., 
1995, ApJ, 438, L63 

\bibitem[]{}
Zdziarski A. A., Johnson W. N., Magdziarz P., 1996, MNRAS, 283, 193 (ZJM96)

\bibitem[]{}
Zdziarski A. A., Lightman A. P., Macio\l ek-Nied\'zwiecki A., 1993, 
ApJ, 414, L93 



\end{thebibliography}
\end{document}